\documentclass[pdflatex,sn-vancouver-ay]{sn-jnl}

\usepackage{graphicx}%
\usepackage{subcaption}%
\usepackage{multirow}%
\usepackage{amsmath,amssymb,amsfonts}%
\usepackage{amsthm}%
\usepackage{mathrsfs}%
\usepackage[title]{appendix}%
\usepackage{xcolor}%
\usepackage{textcomp}%
\usepackage{manyfoot}%
\usepackage{booktabs}%
\usepackage{algorithm}%
\usepackage{algorithmicx}%
\usepackage{algpseudocode}%
\usepackage{listings}%

\usepackage[utf8]{inputenc}
\usepackage[english]{babel}
\usepackage[T1]{fontenc}
\usepackage{textcomp}
\usepackage{lmodern}
\DeclareUnicodeCharacter{00F1}{\~{n}} 
\DeclareUnicodeCharacter{2013}{--}    
\DeclareUnicodeCharacter{2014}{---}   
\makeatletter
\def\UTFviii@defined#1{%
  \ifx#1\relax
    ?
  \else
    \expandafter#1%
  \fi
}
\makeatother
\usepackage{amsmath}
\usepackage{hyperref}
\usepackage{dsfont}
\usepackage{color,soul}
\usepackage{nth}

\newcommand{\figref}[1]{Figure~\ref{#1}}
\newcommand{\tabref}[1]{Table~\ref{#1}}
\newcommand{\eqnref}[1]{Equation~(\ref{#1})}
\newcommand{\secref}[1]{Section~\ref{#1}}

\usepackage{braket}
\usepackage{tikz}
\usepackage{lipsum}
\usepackage{amsmath}
\usepackage{amsfonts}
\newcommand{\ketbra}[2]{\ket{#1}\bra{#2}}

\usepackage{booktabs,tabularx} 

\theoremstyle{thmstyletwo}%

\theoremstyle{thmstylethree}%

\begin{document}

\title[Advantages of Global Entanglement-Distillation Policies in Quantum Repeater Chains]{Advantages of Global Entanglement-Distillation Policies in Quantum Repeater Chains}

\author{\fnm{Iftach} \sur{Yakar}}\email{iftach.yakar@mail.huji.ac.il}

\author{\fnm{Michael} \sur{Ben-Or}}\email{benor@cs.huji.ac.il}

\affil{\orgdiv{School of Computer Science \& Engineering}, \orgname{The Hebrew University of Jerusalem}, \orgaddress{ \city{Jerusalem}, \postcode{91904},  \country{Israel}}}

\abstract{Quantum repeaters are essential for achieving long-distance quantum communication due to photon loss, which grows exponentially with the channel distance.
Current quantum repeater generations use entanglement distillation protocols, where the decision of when to perform distillation depends on either local or global knowledge. Recent approaches for quantum repeaters, such as Mantri et al.
\cite{mantri_comparing_2024}, consider using deterministic local decision policies for entanglement distillation. We ask whether global deterministic policies outperform local ones in terms of communication rate. We simulate equidistant repeater chains, assisted by two-way classical communication, and compare local and global policies for distillation decisions, spanning large distances and varying network and hardware parameters.
Our findings show that global deterministic policies consistently outperform these local ones, and in some cases, determine whether secret communication is possible.
For large repeater chains ($N>512$), global policies improve SKR by two orders of magnitude.
These results suggest that local distillation decisions in quantum repeater chains may not be optimal, and may inform future protocol design.}


\maketitle

\section{Introduction}\label{intro}

Quantum communication holds great promise for secure communication and distributed quantum computing.
However, realizing quantum communication for long distances faces fundamental challenges, owing to the noisy nature of quantum systems and, in particular, the exponential loss of photons over quantum channels, whether through free space or optical fibers.
In contrast with classical information, which can be amplified, quantum information cannot be cloned \cite{wootters_single_1982}, which makes quantum communication over long distances even more difficult. Fortunately, instead of directly sending quantum information over a noisy channel, it is possible to use that channel first to distribute entanglement and then use that entanglement to transfer the information using quantum teleportation.\\

Quantum repeaters are a leading candidate for solving the problem of long-distance quantum communication. Placing repeaters in between the end-to-end parties effectively splits the entire communication length into shorter segments that are less noisy. Along with quantum teleportation and entanglement swapping \cite{zukowski_event-ready-detectors_1993}, repeaters can assist in achieving longer communication lengths \cite{dur_quantum_1999}.
The idea is to use networks, or chains, of intermediate nodes in order to distribute high-fidelity entangled (EPR) pairs, instead of attempting to distribute them directly. Using this approach, the exponential scaling of the fidelity in the quantum channel becomes a polynomial one \cite{muralidharan_optimal_2016}.\\

A key notion in current (\nth{1} and \nth{2}) generations of quantum repeaters is entanglement distillation (also known as entanglement purification), which is a process that converts multiple noisy EPR pairs into fewer, higher-quality pairs.
Due to the high resource cost of distillation, the strategic timing and frequency of such an operation throughout a communication protocol has a significant impact on the overall performance of the repeater chain.
Current approaches often employ local decision rules for distillation, where each repeater can decide whether to perform distillation based on local knowledge it has, such as estimated fidelity and secret key rate ($SKR$) \cite{mantri_comparing_2024}. Furthermore, a distillation decision that depends on real-time information is considered an adaptive decision; otherwise, it is deterministic.\\

While local-knowledge distillation policies have the upshot of reduced communication overhead, they may not achieve the optimal performance across the entire chain or network.
The main question we ask in this work is: \textit{Can deterministic global-knowledge based distillation policies significantly outperform local-knowledge based ones in a repeater chain?}\\

In this work, we study this question by simulating equidistant repeater chains using two-way classical communication with both local and global deterministic distillation policies.
Our contributions are threefold: First, we compare local and global distillation policies across multiple network and hardware parameters.
Secondly, we show that global policies can significantly improve $SKR$ compared to current local policies.
Thirdly, we provide insights on distillation policies in general, and identify parameter regimes where global distillation decisions provide significant advantage.\\

Our results suggest that deterministic local-knowledge based distillation decisions may be sub-optimal across a wide range of parameters. Our simulations show that global policies consistently outperform local ones. In some cases the advantage determines whether any $SKR$ is achievable at all.\\

The paper is organized as follows: \secref{background} provides the necessary background on entanglement purification protocols and quantum repeaters.
\secref{distillation_policies} discusses our taxonomy for different distillation policies.
\secref{near_optimal_distillation_schedules} lays out our methods for finding near-optimal global policies.
\secref{results} show our main results.
\secref{discussion} and \secref{conclusion} discuss directions for future work.

\section{Background}\label{background}

\subsection{Entanglement Purification Protocols (EPPs)}

Entanglement distillation (or: entanglement purification) has been an active topic of research since the mid-90's.
Quantum communication involves transmitting qubits between different parties which are spatially separated.
To transmit qubits, one can encode them on photons in some degree of freedom, e.g., using the photon's polarization or using time-bins \cite{brendel_pulsed_1999}.
Unfortunately, when qubits are sent through an optical fiber or the air (via laser), they are subject to noise.
A prominent noise source is photon-loss, where a sent photon does not reach its destination.
One way to prevent noise from disturbing quantum data is the use of Quantum Error Correction Codes (QECC) \cite{shor_scheme_1995, steane_error_1996-1}.
In QECC, qubits are encoded into larger quantum systems. Using this redundancy in a clever way protects the data against a number of errors, which depends on the code itself.\\

When considering a scenario of communication, however, simply encoding our qubits, sending them over the noisy channel, and then decoding them at the destination may not be ideal.
This is because the encoded information can be lost over the noisy channel.
Assuming good quantum memory is available, the problem of transmitting quantum information can be reduced to creating shared entanglement between two parties and then using quantum teleportation \cite{bennett_teleporting_1993} for sending the information qubits.
Transmitting $n$ arbitrary qubits via teleportation requires $n$ shared EPR pairs between Alice and Bob.
Importantly, in this suggested protocol, the actual information is never transmitted over the noisy channel, only halves of EPR pairs - whose occasional loss is tolerable, as long as their generation rate exceeds the decoherence rate of the memories at each station.\\

Succesful teleportation requires that the EPR pair has good fidelity, and this is where entanglement purification comes in.
Entanglement purification turns $n$ noisy EPR pairs into $k<n$ EPR pairs with superior fidelity. 
Consequently, by incorporating Entanglement Purification Protocols (EPP), the complete teleportation-based communication scheme can be executed even when the channel is noisy, as long as some conditions hold.\\

\subsubsection{Recurrence protocols}\label{recurrence_protocols}

Pioneering works on EPPs include the BBPSSW protocol \cite{bennett_purification_1996} and the DEJMPS protocol \cite{deutsch_quantum_1996}.
These protocols are types of recurrence protocols.
Using these protocols, two parties who share $n$ noisy entangled states can \textit{distill} $k$ maximally entangled states with arbitrary fidelity, where $k<n$.
In other words, by applying the distillation protocol, they turn many previously shared EPR pairs with low fidelity into fewer EPR pairs with higher fidelity.\\

The BBPSSW protocol is as follows: our starting point is Alice and Bob sharing 2 copies of a state $\rho$, where the fidelity between $\rho$ and some maximally mixed state satisfies $F>\frac{1}{2}$.
Typically, these states are achieved by Alice creating an EPR pair and sending half of it to Bob via some noisy quantum channel.
Next, both Alice and Bob perform a set of local operations (Pauli twirling and ) on each of their states, in order to depolarize them and turn them into Werner states.
They then apply $CNOT$ between both of their qubits (this is known as bilateral-CNOT).
Then, both parties measure their second qubit in the computational basis.
Finally, Alice and Bob communicate over a classical channel (this requires two-way classical communication) and compare their measurement results.
If the results disagree, they both throw away the control qubits.
However, if their results agree, they keep the remaining pair which now has higher fidelity which is on average given by:
\begin{align}
  F'=\frac{F^2+[(1-F)/3]^2}{F^2+2F(1-F)/3+5[(1-F)/3]^2}
\end{align}
Where $F$ is the input fidelity, i.e. the fidelity of the states before running this protocol, and $F'$ is the output fidelity.
We have described one step of entanglement distillation.
Recurrence protocols apply this distillation step recursively, on EPR pairs that went through the same number of distillation levels.
When one such step succeeds, the corresponding noisy EPR pair is promoted to the next level, and can then participate in another distillation step together with a pair that has gone through the same number of distillation steps and thus has similar fidelity.
Doing this enough times can achieve arbitrarily high fidelity if the local operations are noiseless.
When a step fails, the corresponding EPR pair is discarded by both Alice and Bob.
In essence, each purification step cuts the total number of EPR pairs in half, while increasing their fidelity exponentially.
This process is expensive in terms of qubits lost, however, it can withstand relatively low initial fidelity ($F>\frac{1}{2}$).\\

A more efficient protocol is the DEJMPS protocol, which is similar to the BBPSSW protocol, but acts on Bell diagonal states instead of Werner states.\\
Dür et al \cite{dur_entanglement_2007} provide a more in-depth description of recurrence protocols.

\subsubsection{Code-based EPP}
There is a close connection between entanglement distillation and error correction which was shown by Bennett et al.\cite{bennett_mixed_1996}.
They showed that entanglement distillation using one-way classical communication is equivalent to error correction in terms of their quantum capacity.
Importantly, they have shown how to construct a one-way entanglement distillation protocol from a general quantum error correcting code.\\

Given a QECC $Q=[[n,k,2t+1]]$ with $n$ physical qubits and $k$ logical qubits, the corresponding entanglement distillation protocol takes in $n$ EPR pairs and turns them into $k$ EPR pairs with higher fidelity.\\

In \cite{aschauer_quantum_2005}, it is shown how to create an entanglement purification protocol (EPP) from a QECC code.
In this protocol, $n$ EPR pairs are shared between Alice and Bob.
As the qubits which are distributed to Bob go through some noisy channel, they might be subjected to noise.
This can be described by:\\
\begin{align}
  E = \mathds{1} \otimes \mathcal{E} \ketbra{\Phi}{\Phi}
\end{align}
As long as the number of qubits affected by noise does not exceed $t$, it is guaranteed that these errors are corrected and $F'=1$ is achieved for the remaining qubits.\\

The protocol can be described as follows:\\
Let $Q$ be a $[[n,k,2t+1]]$ QECC code, where $\mathcal{C}$ and $\mathcal{D}$ are the code's encoder and decoders, respectively.
Alice and Bob start with sharing $n$ noisy EPR pairs.
Alice applies $\mathcal{C}^t$ on her $n$ qubits.
Since $\mathcal{C}$ encodes $k$ logical qubits into $n$ physical qubits, there are $n-k$ "ancilla" qubits.
Similarly, Bob applies $\mathcal{D}$ on his share of the $n$ EPR pairs.
Both Alice and Bob now measure their $n-k$ "ancilla" qubits in the computational basis.
Alice communicates her results to Bob, who then compares his outcomes to hers and applies a correction on the $k$ remaining qubits accordingly.
This can be taken even further, from a deterministic protocol, to a probabilistic one.
If two-way classical communication is allowed, then Bob can also decide to completely discard the remaining $k$ qubits, in case he detects that there are more than $t$ errors.
In this case, Bob can communicate to Alice that she should also discard her corresponding EPR halves.
While an error correction code allows for correcting $t$ errors, it can also be used for error detection and detect up to $d=2t+1$ errors.\\

It is important to note a key practical advantage that EPP has over standard QECC.
In QECC, an arbitrary quantum state is typically encoded, sent through a noisy channel (whether spatially or temporally), error corrected, and decoded.
When the encoded state suffers too many errors passing through the noisy channel, the information is lost.
This poses significant challenges in a computational scenario, for instance.
On the other hand, in EPP, our goal is to distribute EPR pairs that do not encode our information.
Assuming Alice has reliable quantum memory, she can keep her arbitrary, unknown, $n$-qubit state until she and Bob share $n$ high-fidelity EPR pairs, and only then teleport it.
In case one of the stages of entanglement distillation fails, they simply start over, but the information is not lost.

\subsubsection{Combining both methods}
Indeed, distillation using error correction is generally less qubit-consuming than using the recurrence protocols: here $n-k$ EPR pairs are lost, whereas each level of the recurrence protocols uses half of the EPR pairs.
However, for using recurrence protocols, $F>\frac{1}{2}$ suffices, while QECC-based EPP requires the fidelity to be above the error threshold for the code being used.
It is then natural to use both methods in series: as long as the fidelity is above the error threshold, use the recurrence protocols.
Since this increases the fidelity exponentially, no more than a few levels will be required.
Then, once the fidelity is above the error threshold, apply the QECC-based EPP to achieve ideal fidelity with high probability, assuming perfect local operations.

\subsection{Quantum Repeaters}
There are different types of repeaters, canonically categorized into 3 generations \cite{mantri_comparing_2024}:
First-generation quantum repeaters use probabilistic entanglement generation for loss-errors, and heralded entanglement purification for operation errors.
Second-generation quantum repeaters use probabilistic entanglement generation as well, but uses near deterministic error correction for operation errors.
Third generation uses QECC only for both types of errors.
Using current technology, generations $1$ and $2$ are more feasible, but they come with a price: the heralded signals require two-way (classical) communication between checkpoints, which increases the temporal cost and lowers the rate of communication.
Third-generation repeaters, however, usually use only one-way communication, which may potentially yield higher entanglement generation and secret-key rates.\\

In our work, we consider a linear relay network, where the goal is to create EPR pairs between two sides of the relay - Alice and Bob.

\subsubsection{two-way vs one-way repeaters}
In their paper \cite{mantri_comparing_2024}, Mantri et al. propose a two-way protocol for creating entanglement between two stations which are connected via a linear repeater array.
They then compare their protocol to a one-way protocol, in a memory-unconstrained regime, and show that, in their model, the two-way protocol has superior performance in terms of rate and resource cost.
Their protocol, a multiplexed two-way protocol (MTP), is explained in detail in \cite{mantri_comparing_2024}. A concise overview is provided here:\\

\subsubsection{A multiplexed two-way protocol (MTP)}
The MTP is composed of the following steps:
\begin{enumerate}
	\item create initial links (EPR pairs) between neighboring repeaters via intermediate Bell State Analyzer array stations.
	\item recursively apply entanglement purification (EPP) and entanglement swapping until eventually achieving some end-to-end EPR pairs.
\end {enumerate}

The first phase starts by creating multiple EPR pairs between each of the $N$ repeaters.
Spatial or frequency multiplexing enables creation of multiple initial links between each pair of repeaters.
The number of simultaneous link attempts is determined by a parameter $M$ which is typically chosen to be between $128$-$1024$ and assumed to be a power of $2$ for simplicity.
In each segment, midway between neighboring repeaters, a Bell State Analyzer (BSA) array is stationed.\\

Every repeater sends $M$ photons to its neighboring BSA arrays, with the edge stations ($R_0$ and $R_N$) only sending photons in one direction.
The analyzers then perform Bell State Measurement on each of the $M$ pairs, with some probability of creating a link.
The success probability of creating each link between two repeaters is given in \eqnref{eq:link_success_probability}
\begin{equation}\label{eq:link_success_probability}
  \pi_0 = \frac{1}{2} \eta_c^2 e^{-L_0 / L_{att}}
\end{equation}
where $\eta_c$ is the BSA's coupling efficiency, $L_0$ is the distance between the repeaters, and $L_{att}$ is the attenuation length, which is chosen to be $20$ km in \cite{mantri_comparing_2024} and in this work.\\
After this phase, the initial fidelity of each EPR is determined. Again we follow \cite{mantri_comparing_2024} and choose it to be $1- 1.25\epsilon_G$, where $\epsilon_G$ is the gate error rate.\\

Next, entanglement distillation and entanglement swapping is recursively applied between stations in order to achieve high-fidelity end-to-end links.
In each level of the recursion, repeater pairs that share EPR pairs perform optional entanglement distillation, followed by entanglement swapping - to create links across more distant segments.
The model in \cite{mantri_comparing_2024} allows for an optional 1 level distillation in each level, followed by entanglement swapping.
The choice of whether to perform entanglement distillation is based on one of two preselected decision-rules: $SKR$-rule, and $F_{th}$-rule.\\

When using the $SKR$-rule, distillation is performed in step $j$ if it would increase the $SKR$ for that step. Following the $F_{th}$-rule, distillation is done whenever the fidelity $F$ of the EPR-pair(s) is lower than some threshold $F_{th}$.
Assuming $N = 2^n$ segments, the recursion has $n$ levels, after which end-to-end links are created and the final $SKR$ is calculated.

\subsubsection{Upper bounds for linear-repeater chains}
Without any repeaters, the rate of communication is upper bounded by the fundamental repeaterless bound - the PLOB bound \cite{pirandola_fundamental_2017}. An optical loss channel with transmissivity $\eta$ gives a maximum capacity (ebits per protocol use) of:
\begin{equation}
C_{direct} = - \log_2{1 - \eta}
\end{equation}
The basic model of linear repeater chains has been studied before in \cite{pirandola_end--end_2019, guha_rate-loss_2015, azuma_fundamental_2016}.
In \cite{chehimi_scaling_2023}, the authors optimize the number of link-level (level 0) and end-to-end distillation steps.
However, our near-optimal results show that under some parameter regimes, it is better to distill midway, and not right at the link level.
Using first-generation repeaters reduces the exponential scaling of quantum communication to a polynomial one \cite{muralidharan_optimal_2016}.\\

\begin{figure}[!htbp]
\centering
\includegraphics[width=0.85\textwidth]{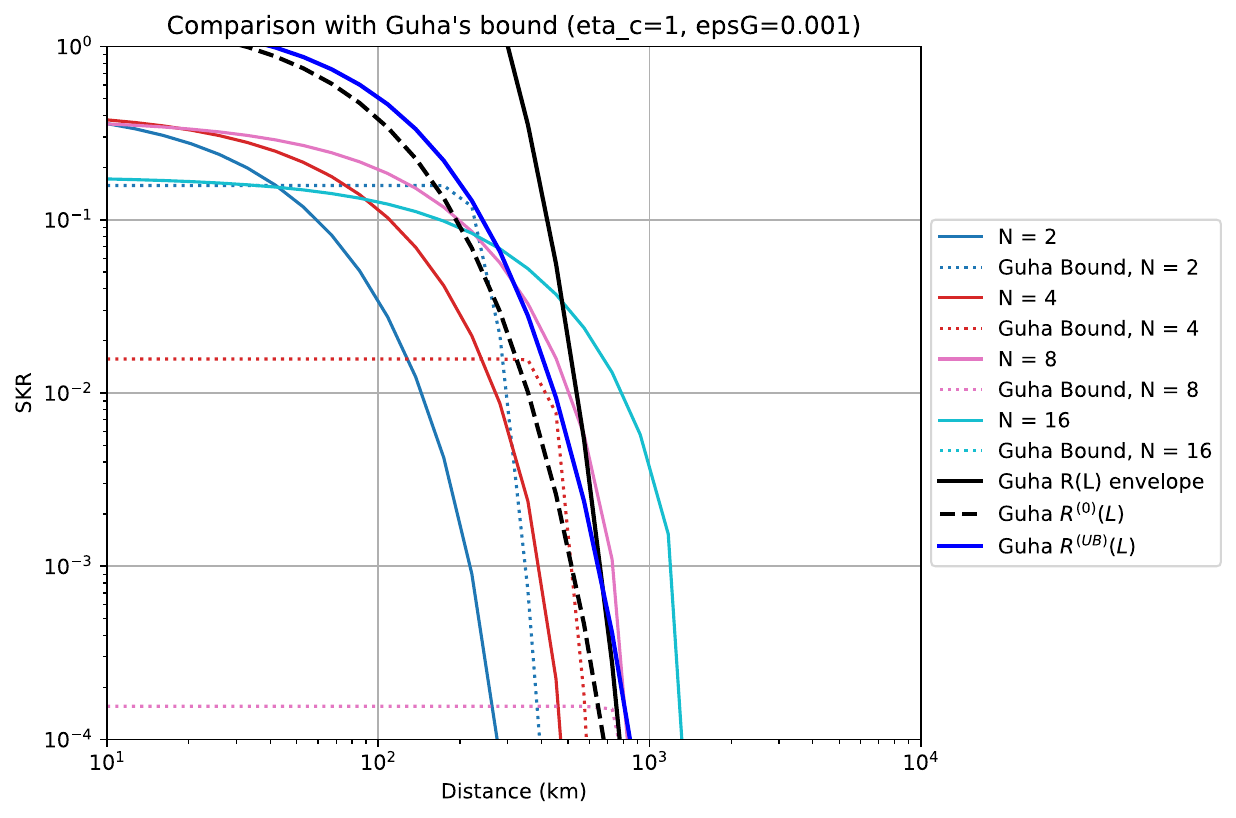}
\caption{Secret Key Rate (SKR) vs. Distance for $\eta = 1$ and $\epsilon = 0.001$, comparing \cite{mantri_comparing_2024} (solid colored curves for different $N$ values) against the bounds from \cite{guha_rate-loss_2015}. Dotted color-coded curves are the three-piece upper bounds $R_N^{\rm UB}(L)$ for each $N$, where $R^{(UB)}(L)$ is their envelope. $R(L)$ and $R^{(0)}(L)$ are rate-loss envelopes as well. For complete definitions see \cite{guha_rate-loss_2015}. }
\label{fig_skr_vs_distance}
\end{figure}

Optimizing swapping policies was studied in \cite{inesta_optimal_2023}, but distillation was not considered in their work.
Guha et al. \cite{guha_rate-loss_2015} present an upper bound for linear repeater chains with multiplexed channel.
However, the model they consider uses only one successful elementary link per burst.
Accounting for more than one successful elementary link requires memory in each repeater and can generally exceed their bound, even for small $N$ such as $N=8$, as shown in \figref{fig_skr_vs_distance}. \\

Adding more nodes to the repeater chain generally increases the upper bound for achievable rates, since each segment can be made less lossy.
Under realistic hardware, however, it is not always the case that adding more repeaters helps.
It was shown in \cite{laurenza_rate_2022} that the rate of such protocols cannot exceed a quantity that only depends on the loss of a single station.\\

A general upper bound, presented in \cite{pirandola_end--end_2019} is referred to as an ultimate bound for repeater chains.
Their bound considers Alice and Bob, with $N$ repeaters between them on a linear chain.
Quantum communication can be in any direction, and classical communication is two-way.
Each quantum channel is used only once.\\
These assumptions capture our model.
In fact, they also capture the \nth{2} and \nth{3} generation of quantum repeaters (QRs).
The ultimate bound for an equidistant $N$ repeater chain, with $\eta$ as the transmissivity of each segment, yields channel capacity:
\begin{equation}
    C_{\text{loss}} (\eta, N) = -\log_2 \left( 1 - \sqrt[N+1]{\eta} \right)
\end{equation}

Note that while this bound in given in terms of channel capacity, it still serves as an upper bound for the SKR.

\section{Distillation Policies}\label{distillation_policies}
In linear repeater chains, deciding when to perform entanglement distillation has a significant effect on the outcomes, in terms of $SKR$.
Distilling too early has the effect of "wasting" too many EPR pairs, with less-than-ideal improvement for that round of distillation.
Conversely, distilling too late generally results in performing distillation with lower input fidelity, and has the risk of fidelity dropping below the threshold fidelity required for distillation $F \geq \frac{1}{2}$.\\

One way of categorizing distillation policies is based on their local/global and deterministic/adaptive properties. This gives us four classes: Local-deterministic (LD), Global-deterministic (GD), Local-adaptive (LA), and Global-adaptive (GA). Informally, deterministic policies don't depend on real-time information, whereas adaptive policies may do so. Defining Local policies is more subtle in the deterministic regime, but can be thought of as a node that decides on distillation without considering global information.

\subsection{Local-deterministic (LD)}
In local-deterministic policies, each repeater can decide whether to perform distillation or not, based on some rule that is not dependent on the repeater's location in the chain, or the distillation level it belongs to.
These policies are deterministic since they only use information which can be pre-calculated (e.g., probability distribution of number of links between stations, and expected fidelity), and never look at real-time data for deciding on distillation.\\

Such policies are used in \cite{mantri_comparing_2024} and are referred to as the $F_{th}$-rule and the $SKR$-rule.
The $F_{th}$-rule depends on the expected pair average fidelity at each step, where the $SKR$-rule depends on the $SKR$ at each step, which can be pre-calculated based on the expected probability distribution of the number of links and the average fidelity.
Note that in this model, every repeater participates in a single recursion step, and so we can say that the policy is "stepwise local".

\subsection{Global-deterministic (GD)}
These policies consider the entire protocol in an attempt to optimize some metric, such as $SKR$. They are still deterministic and don't consider any mid-run information. One advantage of using GD policies over GA policies, is that repeaters never have to wait for any information they might be missing in order to decide on distillation. Since idle qubits are subject to memory decoherence, any additional time spent waiting degrades the fidelity.\\

It is natural to wonder what exactly distinguishes local-deterministic policies and global-deterministic ones, if one was to rigorously define it.
One approach would be to require that distillation rules under local-deterministic policies are not dependent on the protocol level.
The $F_{th}$ rule trivially satisfies that constraint since it is only a function of the current fidelity at each node.\\

The $SKR$ rule, however, might be more delicate, since it is dependent on the current SKR, the calculated distribution of links for that stage and the expected average link state.
The input here may be rich enough that one could claim that it easily encodes the protocol level, even if not explicitly. Nevertheless, for our purposes, we consider both $F_{th}$ and $SKR$ rules to belong to the $LD$ class.

\subsection{Adaptive Policies}

Policies that consider real-time information which is not known in advance are considered adaptive policies. Here, again, one can consider local and global policies.

\begin{enumerate}
    \item \textbf{Local-adaptive} - A repeater that considers local real-time data to decide on distillation. For instance, the number of successful links between a repeater with its neighbors, or measured fidelity can potentially help a repeater make better decisions.

    \item \textbf{Global-adaptive} - In global-adaptive policies, the entire state of the repeater chain can be considered in real-time. This has the advantage of being able to use more knowledge than other classes, but, contrary to deterministic policies, may increase wait times, which in general decrease the output $SKR$ value for two reasons:
    \begin{enumerate}
        \item Additional wait-time exposes the memory qubits to noise for longer times, which introduces errors.
        \item Increasing the time spent on each level of the protocol increases the time needed to create end-to-end links, thus impacting SKR when measured in time. Note that in this work SKR is defined per channel use, and not per unit time.
    \end{enumerate}
    In \cite{inesta_optimal_2023}, the authors study global-adaptive policies for entanglement-swapping decisions.
\end{enumerate}

\subsubsection{Possible Improvements by Adaptive Policies}
A natural question to ask is how much one can improve this protocol by allowing the distillation decision to be decided in real time?
More specifically, one can ask the same question for the discussed distillation rules, $F_{th}$ and $SKR$, when relying on adaptive knowledge. While we don't consider adaptive policies in our simulations, we do provide some relevant insights in this section.\\

It would seem that the $SKR$-rule decision depends on the successful links, and that knowing the actual number of successes might provide an advantage. However, note that this information does not affect the decision, since the number of successful links appears both in the pre/post-expressions and cancels out, and so there's no advantage there.
Regardless, knowing the number of links in real-time is trivial (and free) since the BSAs generate a heralding signal that informs the repeaters which links were successful.\\

In contrast, estimating the fidelity using real-time is less trivial.
In an average scenario, the average fidelity can be calculated in each step of the protocol ahead of time.
Finding a better estimation for fidelity based on real-time information is a bit more involved. It is possible, of course, to sacrifice some of the EPR pairs by measuring them and calculating the expected fidelity.
Unfortunately, while in asymptotic regimes it is possible to measure a constant number of qubits to estimate the true fidelity with small error, in current physical systems, it is relatively expensive in terms of EPR pairs sacrificed.
This is especially prominent in the low-error regime, like ours.
For example, even measuring $10\%$ out of $128$ links, when the initial fidelity is $0.99$, will not be more accurate than what can be calculated a priori.\\ 

One possible approach could be to keep the information on entanglement distillation/swapping success for the next steps. In particular, since the distillation success rate depends on the input fidelity, knowing how many pairs distilled successfully, can inform us of the average input fidelity of these pairs, without sacrificing any of them. One caveat for this approach is that the adaptive knowledge associated with it is only learnt after the first distillation. In other words, the first decision cannot depend on such information. 

\subsection{Alternative distillation categorization}

There can be, of course, other categories for policies.
For example, \cite{haldar_reducing_2024} consider a model with quasi-local knowledge-based decisions.
Quasi-local here means that every repeater has knowledge of some of its neighbors but not necessarily full global knowledge.
In addition to studying quasi-local policies, they also consider questions on distillation policies, such under what parameter regimes it is better to first swap and then distill (\textbf{SWAP-DISTILL}) or vice-versa (\textbf{DISTILL-SWAP}).
However, they don't consider more than one distillation step, since they always distill $k$ EPR pairs into a single pair.

\section{Methods for finding Near-optimal distillation schedules}\label{near_optimal_distillation_schedules}
In \cite{mantri_comparing_2024}, the authors base their distillation schedule on local decisions.
In this section, we consider optimizing over global distillation schedules.

We show that such global distillation schedules outperform the local distillation rules used in \cite{mantri_comparing_2024}.

\subsection{A global search for good schedules}\label{global_search}
A simple and naive approach to finding better distillation schedules is to randomly sample a vector $\mathbf{D} \in \mathbb{N}^n$ where $D_i$ is the number of distillation steps at level $i$ of the protocol.
Doing this many times, and simulating the protocol for each choice, produces schedules better than the ones produced by the $F_{th}$ and $SKR$ rules.\\

Empirically, it seems that this process indeed yields near-optimal schedules.
Indeed, schedules that are close under a simple distance metric (e.g. L1 on the schedule vector) consistently yield similar SKR.
Moreover, for fixed $N$ and $M$, varying the other parameters produces similar best schedules, while being independently sampled.
This suggests that our near-optimal schedules are near the true optimum, though it is only a heuristic. \\

Notice that, in contrast with \cite{mantri_comparing_2024}, we allow for more than 1 distillation step per protocol level.
In addition, we allow for distillation to happen right after creating the initial link, as well as after establishing the end-to-end link.
Here, and across all of our simulations, only the DEJMPS EPP is considered.
The search space can be reduced by observing that since the DEJMPS protocol sacrifices half of the qubits for each step, the total distillation budget is $\sum{D_i} \leq \log_2{M}$, after which no further distillation can be performed.
Random sampling of such vectors of $D$, under these constraints, yields distillation schedules that exceed both the $SKR$ local distillation rule, and the $F_{th}$ rule under different $F_{th}$ values.\\

We highlight that using this Monte Carlo approach to find a near-optimal policy is costly in terms of computation. While this search happens offline, before running the protocol, reducing the runtime is nonetheless relevant in a real-world scenario, since it impacts the total communication time, even if not affecting the $SKR$.\\

Reducing the offline calculation time is certainly possible. For example, in our setting, we have not bounded the number of allowed distillation steps in each protocol level.
This makes the search space much larger than it needs to be, since using too many distillation steps is inefficient.
In our simulations, we have found that none of the near-optimal policies use more than 2 distillation steps per level. This suggests that putting an upper bound of 2-3 steps per level might not prune any optimal policy.
Moreover, there are far better tools for finding optimal solutions than Monte Carlo, like genetic algorithms, Bayesian optimization, or Markov Decision Process, as used in \cite{inesta_optimal_2023}.

\subsection{Sampling methodology}
In order to find near-optimal policies, we used the Monte-Carlo process described in \secref{global_search} with different parameters. Our parameter space is specified in \tabref{tab:parameters_table}.
For every combination of parameters, we ran the simulation 500 times, each time with a different random choice of the distillation schedule $\mathbf{D}$, and selected the one that achieved the highest $SKR$ as the near-optimal policy.
For all of the chosen policies, we find that they use at most 2 distillation steps per protocol level.

\begin{table}[h]
    \small
    \setlength{\tabcolsep}{4pt}
    \renewcommand{\arraystretch}{1.15}
    \begin{tabularx}{\linewidth}{@{}l p{0.32\linewidth} X@{}}
        \toprule
        \textbf{Parameter} & \textbf{Allowed values} & \textbf{Definition} \\ \midrule
        $N$              & $\{\,2^{k}\mid k = 2,\dots,12\,\}$ & Number of repeater segments in the chain \\[2pt]
        $M$              & $\{512,\,1024,\,2048\}$                                            & Number of multiplexed elementary links \\[2pt]
        $\eta_c$         & $\{0.3,\,0.5,\,0.9,\,1.0\}$                                        & Bell‐state‐analyzer (BSA) coupling efficiency \\[2pt]
        $\epsilon_G$     & $\{10^{-4},\,10^{-3}\}$                                            & Two‐qubit gate error probability per operation \\[2pt]
        Total distance   & $[10,\,10^{4}]\,\mathrm{m}$                                        & End‐to‐end physical distance of the repeater chain \\ \bottomrule
    \end{tabularx}
    \caption{Parameter space used in the simulations. As for $\eta_c$ and $\epsilon_G$, we consider the values used in \cite{mantri_comparing_2024}. $\eta_c$ values $\{0.3,\,0.5,\,0.9,\,1.0\}$ are denoted "low coupling", "medium coupling", "high coupling", and "perfect coupling", respectively. $\epsilon_G$ values $\{10^{-4},\,10^{-3}\}$ are denoted "low gate errors" and "moderate gate errors", respectively.}
    \label{tab:parameters_table}
\end{table}

\subsection{Simulating the protocol}
Calculating the SKR for a given set of parameters is done following the MTP protocol described in \cite{mantri_comparing_2024}.
All analytical expessions used in these calculations are taken from Appendix A of \cite{mantri_comparing_2024}.\\

The steps to simulate the protocol are given in \ref{alg:mtp}. Each run produces a single deterministic $SKR$ result for the given parameters.
In the sampling process, \ref{alg:mtp} is used for evaluate $SKR$ for a chosen $\mathcal{D}$ by selecting the $\texttt{MANUAL}$ policy.

\begin{algorithm}[htbp]
  \caption{multiplexed two-way protocol (MTP) Simulation (single distance)}
  \label{alg:mtp}
  \begin{algorithmic}[1]
  \Require
    $N, M, \epsilon_G, \eta_c, L_{\mathrm{tot}}, 
     \mathcal{R}$ \Comment{$\mathcal{R}\in\{$\texttt{F\_th}, \texttt{SKR}, \texttt{MANUAL}$\}$}
  \Require
    Optional schedule $\mathbf{D}$ (\texttt{MANUAL} mode)
  \Require
    Optional threshold $F_{th}$ (\texttt{F\_th} mode)
  \Ensure Secret-key rate $\mathrm{SKR}$ and executed schedule $\mathbf{D}$
  
  \State $L_0 \gets L_{\mathrm{tot}}/N$ \Comment{segment length}
  \State $\pi_0 \gets \textsc{LinkSuccessProbability}(\eta_c, L_0)$ \Comment{\eqnref{eq:link_success_probability}}
  \State $p \gets \operatorname{Binomial}(M,\pi_0)$ \Comment{link-count distribution}
  \State $\rho \gets \textsc{ElementaryLinkInit}(\epsilon_G)$ \Comment{Initial average state}
  \State $L \gets \lfloor\log_2 N\rfloor$ \Comment{number of repeater levels}
  
  \For{$i \gets 0$ \textbf{to} $L-1$} \Comment{iterate protocol levels}
    \State $d \gets
      \begin{cases}
         \mathbf{D}[i] & \mathcal{R}=\texttt{MANUAL}\\
         f_{\mathcal{R}}(p,\rho) & \text{otherwise}
      \end{cases}$ \Comment{how many distillation steps}
    \If{$d>0$}
       \State $(p,\rho)\gets\textsc{Distill}(p,\rho,d)$
    \EndIf
    \State $(p,\rho)\gets\textsc{Swap}(p,\rho)$ \Comment{entanglement swapping}
    \State $\rho\gets\textsc{Dephase}\bigl(\rho,\,L_0 2^{i}\bigr)$ \Comment{memory decoherence}
  \EndFor
  
  \State $\mathrm{SKR}\gets\textsc{ComputeSKR}(p,\rho,M)$
  \State \Return $(\mathrm{SKR},\mathbf{D})$
  \end{algorithmic}
  \end{algorithm}

\section{Results}\label{results}
Since performing even a single additional distillation step reduces the number of EPR pairs by half, deciding when to distill is crucial for optimizing key rates.
Under some parameter regimes, this decision can determine whether the SKR is nonzero or vanishes.\\

Viewing all the results collected (policy comparison, 3D plots, plateau-ratio plots) across different parameter configurations is possible using this interactive web-app: \cite{global_epp_results}.

\subsection{Effect of gate error rate}
\begin{figure}[!htbp]
    \centering
    \begin{subfigure}{0.45\textwidth}
      \centering
      \includegraphics[width=\textwidth]{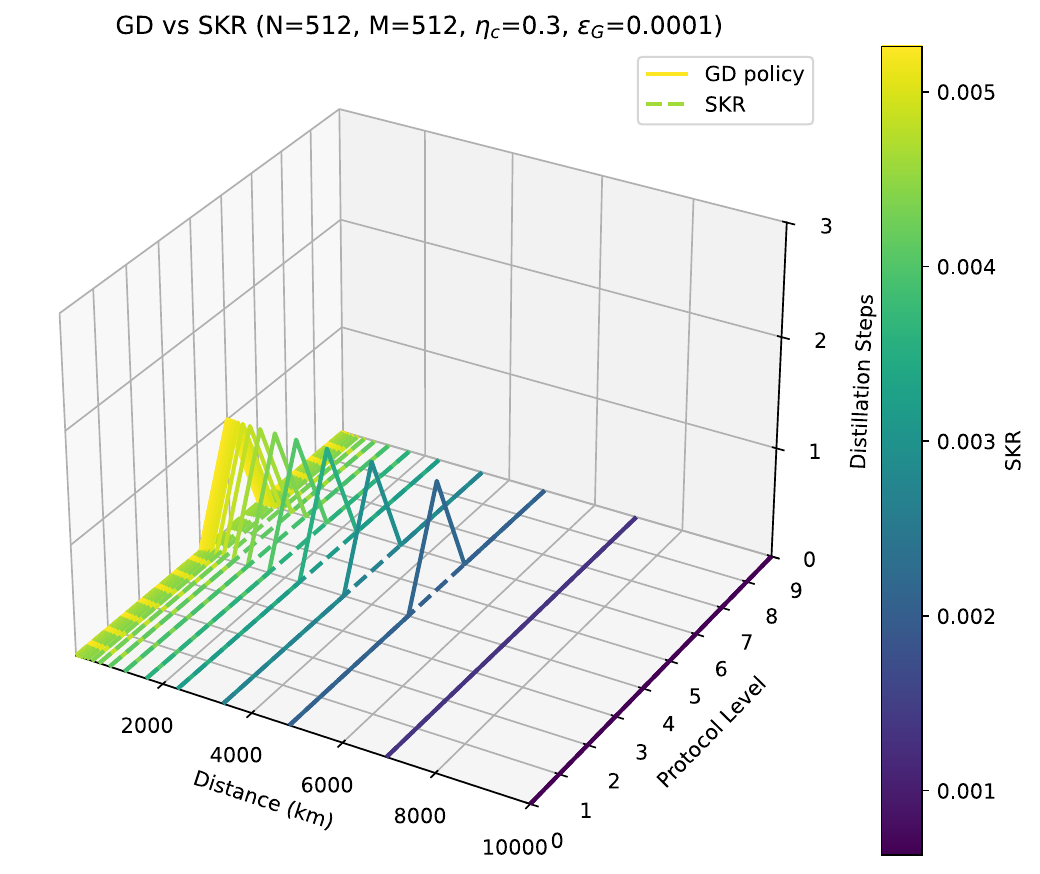}
      \caption{$\epsilon_G = 0.0001$}
      \label{fig:3d_visualization_epsg_0.0001}
  \end{subfigure}
    \hfill
    \begin{subfigure}{0.45\textwidth}
      \centering
      \includegraphics[width=\textwidth]{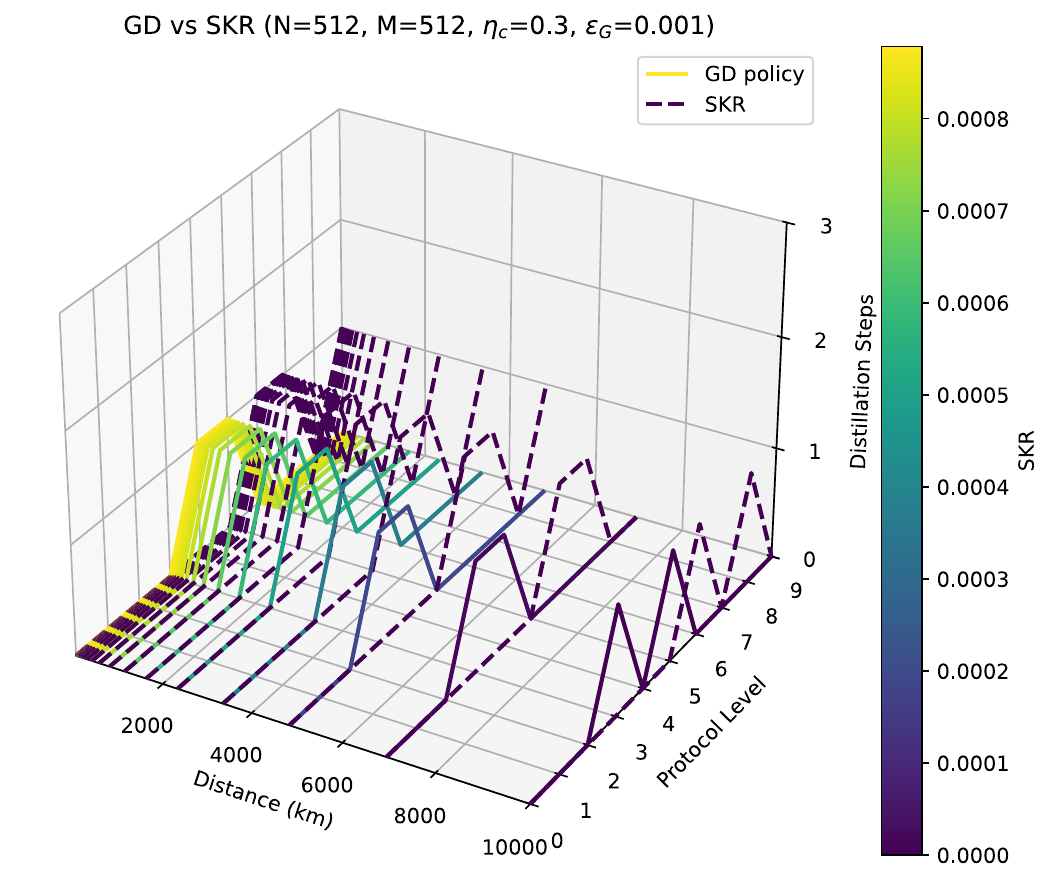}
      \caption{$\epsilon_G = 0.001$}
      \label{fig:3d_visualization_epsg_0.001}
  \end{subfigure}
    \caption{Comparisons between GD policy and LD policy (SKR rule) for $N,M=512, \eta_c=0.3$. 3D plots show the number of distillation steps at each level of the protocol for varying distances. The plots are color coded by the output SKR. \\
    (a) shows the low error regime ($\epsilon_G = 10^{-4}$) where the GD policy distills once at level 5, and the SKR policy does not distill at all.
    (b) shows the moderate error regime ($\epsilon_G = 10^{-3}$), where the GD policy distills earlier at first, but then does not redundantly distill later on.}
    \label{fig:3d_visualization_comparison}
\end{figure}

As we will see in \secref{plateau_ratio}, the gate error rate has a significant effect on SKR for all policies and, in particular, on the scale of the advantage from using GD policies.
\figref{fig:3d_visualization_comparison} shows a an example comparison of the distillation schedules used by different policies for different gate error rates.
In this example, the presence of moderate errors yields a significant advantage of using the GD policy, by choosing a better schedule.
In contrast, in the low error regime ($\epsilon_G = 10^{-4}$), the advantage for the GD policy is less pronounced.\\

An important observation is that in our model, the initial fidelity is not affected by the distance between two repeaters. Consequently, the $F_{th}$ policy effectively employs the same schedule, independently of the distance.
For the $SKR$-rule, however, the situation is different. The number of successful initial links decreases exponentially with the segment distance.
Since the $SKR$ decreases as $M$ increases (see Appendix A of \cite{mantri_comparing_2024}), this policy yields different $D$'s for different end-to-end distances.

\subsection{Critical advantage for large $N$}\label{critical_advantage_for_large_N}
\begin{figure}[!htbp]
  \centering
  \begin{subfigure}{0.45\textwidth}
    \centering
    \includegraphics[width=\textwidth]{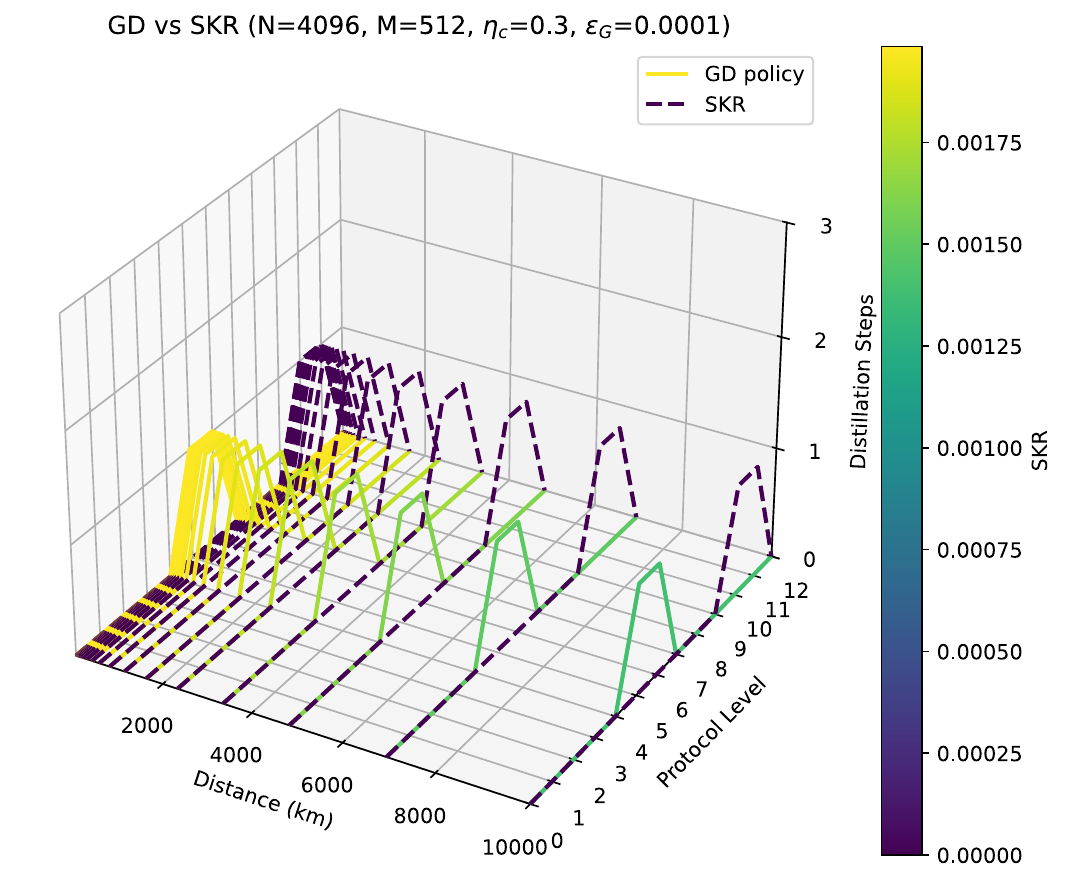}
    \caption{$\eta_c = 0.3$}
    \label{fig:3d_visualization_N4096_etac_0.3}
\end{subfigure}
  \hfill
  \begin{subfigure}{0.45\textwidth}
    \centering
    \includegraphics[width=\textwidth]{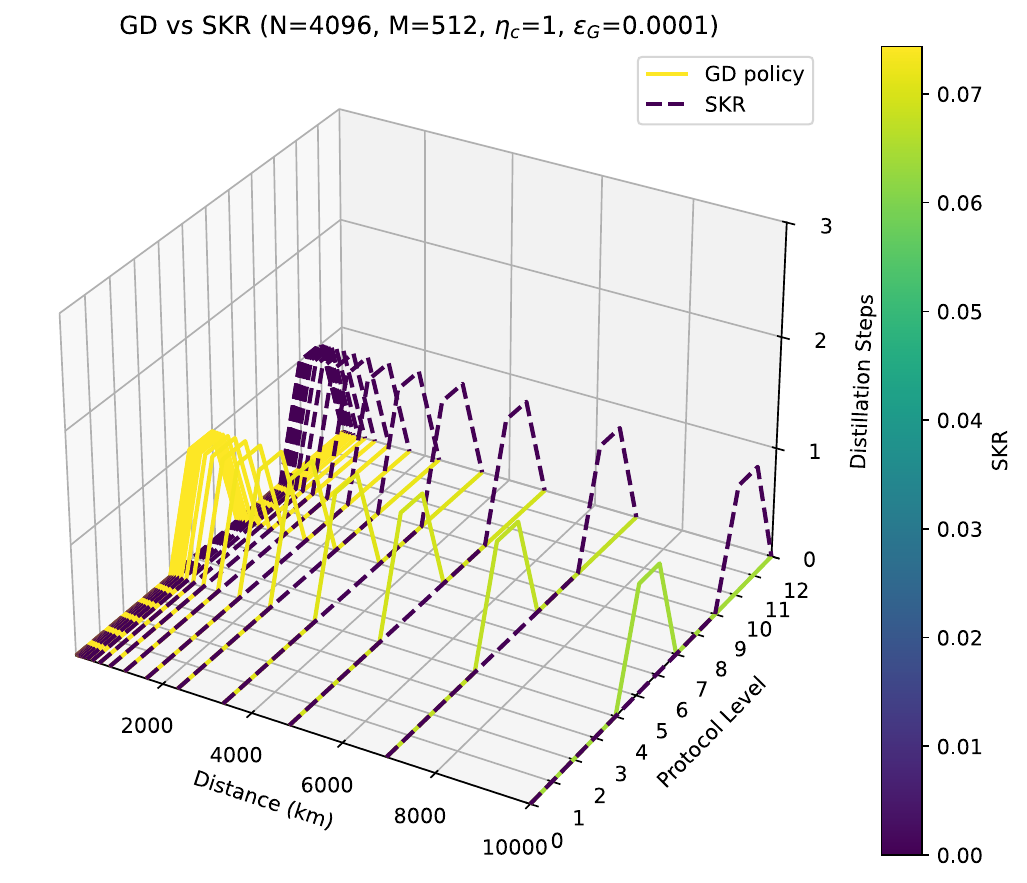}
    \caption{$\eta_c = 1.0$}
    \label{fig:3d_visualization_N4096_etac_1.0}
\end{subfigure}
  \caption{Comparisons between GD policy and LD policy (SKR rule) for $N=4096,M=512, \epsilon_G=0.0001$. \\
  (a) shows the low coupling regime ($\eta_c=0.3$), and (b) shows the perfect coupling regime ($\eta_c=1.0$).}
  \label{fig:3d_visualization_comparison_N4096}
\end{figure}

There are some cases where secret communication is possible only with the GD policy.
For example, \figref{fig:3d_visualization_comparison_N4096} shows how for large $N$ values, GD policy can achieve a non-zero SKR, while the SKR rule cannot.
In this example, the SKR rule distills at the end, but it is too late. The GD policy, however, distills about halfway of the chain, preventing the fidelity from dropping too much, and manages to produce a nontrivial rate.

\subsection{Plateau ratio}\label{plateau_ratio}
When considering SKR vs. distance, the plot typically starts as a plateau and then drops at longer distances, which is characteristic of this setting with any parameter combination.
In order to compare between different policies, and capture the advantage of GD policies, it would be useful if the distance dimension could be ignored.
The average plateau SKR allows us to do that.\\

\begin{figure}[!htbp]
    \centering
    \includegraphics[width=0.85\textwidth]{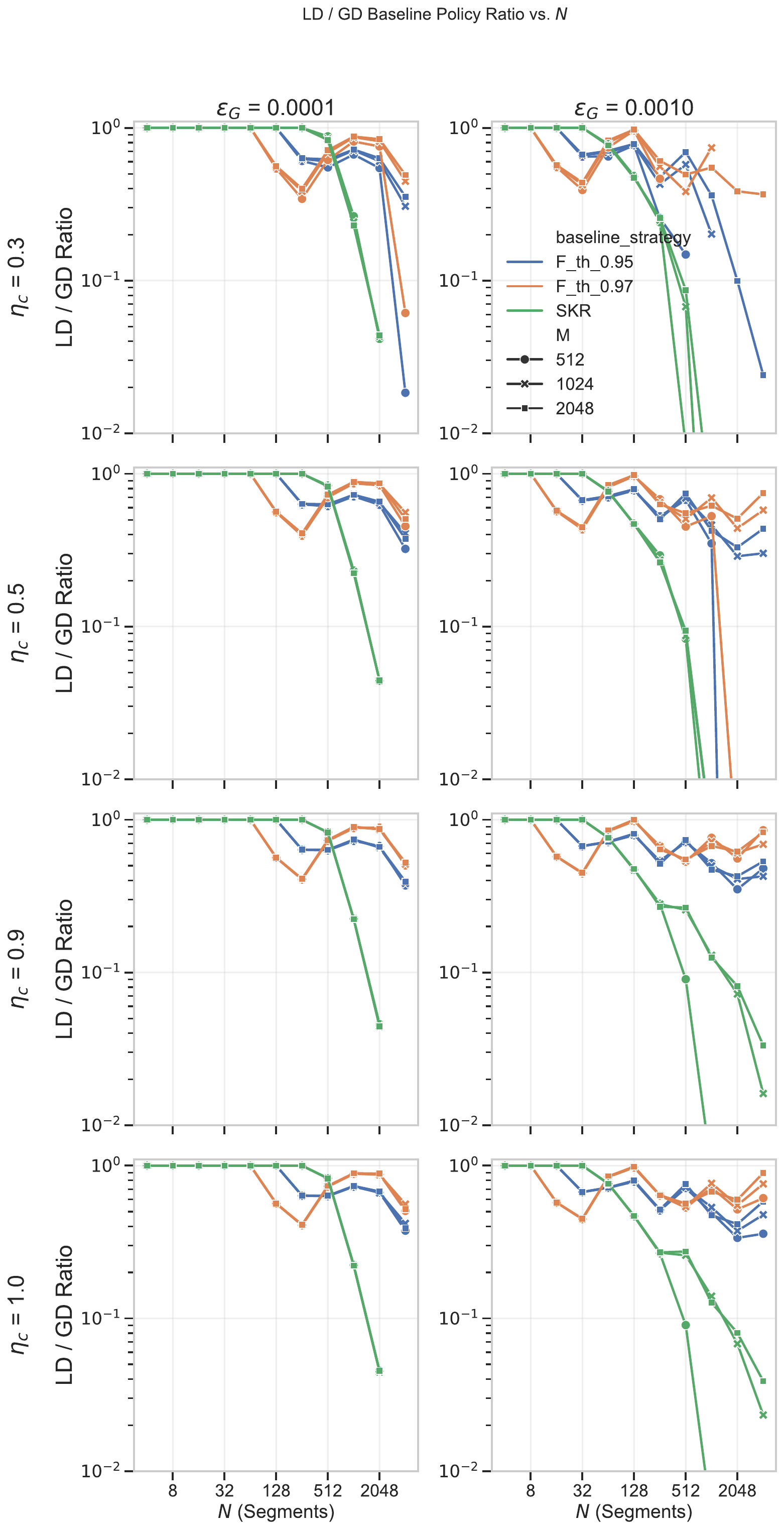}
    \caption{Inverse plateau ratio (LD/GD) vs. number of segments ($N$) for different gate error rates ($\epsilon_G$) and coupling efficiencies ($\eta_c$). Each line represents a different multiplexing value ($M$) or distillation rule. For clarity, we plot the inverse of the plateau ratio $\mathrm{LD}/\mathrm{GD}=\text{plateau ratio}^{-1}$ such that values below 1 indicate an advantage for the GD policy.
    Data points have been omitted in either of these cases: (1) baseline LD policy produces a negligible SKR, yielding an undefined ratio, and (2) both baseline and GD policies produce negligible SKR values.}\label{fig_plateau_ratio}
\end{figure}

\figref{fig_plateau_ratio} shows a comparison between GD and LD policies based on the plateau ratio metric, which we introduce.
First, the maximal SKR value for each parameter combination is identified. The plateau region is then defined as all points where the SKR exceeds 0.9 of the maximal SKR. The average plateau SKR is calculated by taking the mean of SKR values within this region. The plateau ratio is then computed as:

\begin{equation}
    \text{Plateau ratio} = \frac{\text{Average plateau SKR}_{\text{GD}}}{\text{Average plateau SKR}_{\text{LD}}}
\end{equation}

This ratio directly measures the relative advantage of GD over LD policies, with values above 1 indicating GD outperforms the baseline LD policy.
We chose the threshold to be $0.9$, which empirically captures the plateau region while excluding the drop. \\

Looking at these values enables direct comparison between GD and LD policies, and shows how the advantage scales as $N$ and $M$ change.
In particular, we see that in the moderate gate error regime ($\epsilon_G = 10^{-3}$), the plateau ratio can go up to two orders of magnitude for higher $N$ values, while for lower gate errors, the advantage is less pronounced.
Moreover, lower $M$ values also show a more significant advantage in some cases, in particular $M=512$. This is probably since lower $M$ has a lower distillation budget.
As $N$ increases, the advantage over $SKR$ rules increases monotonically, and often much faster than the other policies, while the advantage over $F_{th}$ rules is more variable.
We also observe a pattern where the $F_{th}$ rule underperforms the $SKR$ rule for lower $N$ values, but outperforms it above some $N$.\\

These results suggest that the higher the error rate and the number of repeaters, using GD policies becomes more substantial.

\subsection{Minimal $N$ for advantage}
From a practical perspective, it is natural to ask \textit{how many segments are enough for a global over local advantage?} \figref{fig_plateau_ratio} shows the answers to this question for various network parameters and distillation rules. In general, higher $\epsilon_G$ values favor GD policies starting at lower $N$ values.
E.g., for the $SKR$ rule, $N=64$ segments are enough for a global advantage with moderate gate errors, while with low gate errors at least $512$ segments are needed.
In addition, we see that in the sense of the minimal $N$ for advantage, $F_{th}$ strategies are consistently easier to beat than $SKR$ ones across all the parameters simulated.
These results are consistent across varying values of $M \in [512, 1024, 2048]$ and $\eta_c$.

\section{Discussion}\label{discussion}

Across the parameters we have used, our near-optimal policies consistently outperformed both the $F_{th}$ and the $SKR$ rules. Under some parameter regimes, the improvement was so vast, that it determined whether any secret communication was possible.\\

Intuitively, local rules are often suboptimal. When each level (or node) attempts to optimize its own success, it does not consider the entire chain. The distillation schedules produced by local rules often under- or overspend their budget $\log_2(M)$. Since every distillation operation destroys at least half of the qubits in that level - the timing is crucial. A global policy balances the budget in order to optimize for some end-to-end result. It finds the ideal locations for distillation, which the local approaches might not achieve. The figure of merit in our case is the $SKR$, which combines both the average fidelity and the number of links into one number. However, one can optimize a global policy with respect to other metrics.\\

Protocol designers should precompute a global distillation schedule as a function of the known system parameters (see \tabref{tab:parameters_table}) and build a lookup table for a given chain. Repeaters can then find the near-optimal distillation schedule for different values of $N, M, distance$ .\\

Furthermore, even though we have only considered \nth{1} generation repeaters, the same principles are applicable to other generations. For example, a \nth{3} generation protocol, that error-corrects encoded qubits at each node, might introduce more noise than it corrects under some parameter regimes.\\

There are, of course, some downsides to using global policies: they may require pre-calculations or some centralized control. This trade-off may certainly not be trivial in real systems.\\
 
While we improved on \cite{mantri_comparing_2024}, there is still much work to close the gap with the theoretical ultimate bound from \cite{pirandola_end--end_2019}.

\subsection{Limitations}\label{limitations}

Our work has the following limitations:

\begin{itemize}
    \item \textbf{Considering only DEJMPS protocol}\\
    Our work only considered the DEJMPS protocol.
Code-based EPPs can have improved performance under some parameter regimes \cite{dur_entanglement_2007}, and therefore their incorporation has the potential to increase the performance of such protocols.
Allowing for code-based EPP might change how global policies compare to the local ones. 
We expect that combining code-based EPPs will give an even larger advantage to global policies, since such protocols generally have a more significant effect on EPR count.\\

\item \textbf{Deterministic only}\\
We have not run simulations for adaptive protocols.
It is not clear whether global policies outperform local policies in the adaptive regime.
This is due to the fact that a global adaptive policy must have a significant classical communication overhead that is necessary for real-time information transmission between repeaters.\\

\item \textbf{Considering linear chains}\\
We have not looked at network structures that are more complex than a linear chain. It is not clear whether the global advantage is maintained in such networks. Moreover, finding near-optimal policies in complex network structures is less trivial, and might not scale well with our naive Monte Carlo approach.

\end{itemize}

\section{Conclusion and Future Work}\label{conclusion}

Local policies do not use the entire potential in terms of SKR in the deterministic regime. Our global policies make better use of the given resources.\\

Future work may include:
\begin{itemize}
    \item \textbf{Separating Local and Global} \\
    \textit{Under a more rigorous definition of separation between local and global policies, are LD policies non-optimal in general?}\\

    \item \textbf{Better local distillation rules} \\
    \textit{Are there local distillation rules that yield SKRs that approach those possible with GD?}\\

    \item \textbf{Combining code-based EPPs} \\
    \textit{How does combining code-based EPPs in both LD and GD policies affect the gap between these policies?}
    Note that doing this would necessitate designing more complex local decision rules for the LD policies, as well as better optimization methods for the GD ones.\\
    It can be interesting to consider codes that allow for error detection, like the tesseract code introduced in \cite{reichardt_demonstration_2024}. This might improve fidelity by post-selection.

    \item \textbf{Adaptive policies} \\
    \textit{Do adaptive policies have an advantage over deterministic ones, under practical parameter regimes?}
    These typically require more complicated simulations, possibly involving Monte Carlo sampling.\\

\end{itemize}

This work has laid the groundwork for practical, precomputed distillation schedules in current quantum repeater protocols.

\bmhead{Supplementary information}

An interactive web-app with all our results is available at \cite{global_epp_results}.

\bmhead{Acknowledgements}

We would like to thank David Ponarovsky and Prateek Mantri for the helpful discussions.

\section*{Declarations}

All analytical expressions used for the simulations of the protocol were taken from Appendix A of \cite{mantri_comparing_2024}.\\
Code used for generating the data is available upon reasonable request.\\
There are no conflicts of interest to declare.\\
Funding: Research supported in part by Israel Science Foundation Grant No. 2137/19; Israel Innovation Authority Grant; and by a DDR\&D research grant.\\

\bibliography{distillation, manual}

\begin{thebibliography}{24}
\providecommand{\natexlab}[1]{#1}
\providecommand{\doi}[1]{\url{https://doi.org/#1}}
\providecommand{\url}[1]{\texttt{#1}}
\providecommand{\urlprefix}{}

\bibitem[{Aschauer(2005)Aschauer, Hans}]{aschauer_quantum_2005}
Aschauer H.
\newblock Quantum communication in noisy environments.
\newblock Text.{PhDThesis}, Ludwig-Maximilians-Universität München; 2005.

\bibitem[{Azuma et~al.(2016)Azuma, Koji and Mizutani, Akihiro and Lo, Hoi-Kwong}]{azuma_fundamental_2016}
Azuma K, Mizutani A, Lo HK.
\newblock Fundamental rate-loss trade-off for the quantum internet.
\newblock Nature Communications. 2016 Nov;7(1):13523.
\newblock \urlprefix\url{https://www.nature.com/articles/ncomms13523}, publisher: Nature Publishing Group, \doi{10.1038/ncomms13523}.

\bibitem[{Bennett et~al.(1993)Bennett, Charles H. and Brassard, Gilles and Crépeau, Claude and Jozsa, Richard and Peres, Asher and Wootters, William K.}]{bennett_teleporting_1993}
Bennett CH, Brassard G, Crépeau C, Jozsa R, Peres A, Wootters WK.
\newblock Teleporting an unknown quantum state via dual classical and {Einstein}-{Podolsky}-{Rosen} channels.
\newblock Physical Review Letters. 1993 Mar;70(13):1895--1899.
\newblock \urlprefix\url{https://link.aps.org/doi/10.1103/PhysRevLett.70.1895}, publisher: American Physical Society, \doi{10.1103/PhysRevLett.70.1895}.

\bibitem[{Bennett et~al.(1996{\natexlab{a}})Bennett, Charles H. and Brassard, Gilles and Popescu, Sandu and Schumacher, Benjamin and Smolin, John A. and Wootters, William K.}]{bennett_purification_1996}
Bennett CH, Brassard G, Popescu S, Schumacher B, Smolin JA, Wootters WK.
\newblock Purification of {Noisy} {Entanglement} and {Faithful} {Teleportation} via {Noisy} {Channels}.
\newblock Physical Review Letters. 1996 Jan;76(5):722--725.
\newblock \urlprefix\url{http://arxiv.org/abs/quant-ph/9511027}, arXiv:quant-ph/9511027, \doi{10.1103/PhysRevLett.76.722}.

\bibitem[{Bennett et~al.(1996{\natexlab{b}})Bennett, Charles H. and DiVincenzo, David P. and Smolin, John A. and Wootters, William K.}]{bennett_mixed_1996}
Bennett CH, DiVincenzo DP, Smolin JA, Wootters WK.
\newblock Mixed {State} {Entanglement} and {Quantum} {Error} {Correction}.
\newblock Physical Review A. 1996 Nov;54(5):3824--3851.
\newblock \urlprefix\url{http://arxiv.org/abs/quant-ph/9604024}, arXiv:quant-ph/9604024, \doi{10.1103/PhysRevA.54.3824}.

\bibitem[{Brendel et~al.(1999)Brendel, J. and Gisin, N. and Tittel, W. and Zbinden, H.}]{brendel_pulsed_1999}
Brendel J, Gisin N, Tittel W, Zbinden H.
\newblock Pulsed energy-time entangled twin-photon source for quantum communication.
\newblock Physical Review Letters. 1999 Mar;82(12):2594--2597.
\newblock \urlprefix\url{http://arxiv.org/abs/quant-ph/9809034}, arXiv:quant-ph/9809034, \doi{10.1103/PhysRevLett.82.2594}.

\bibitem[{Chehimi et~al.(2023)Chehimi, Mahdi and Pouryousef, Shahrooz and Panigrahy, Nitish K. and Towsley, Don and Saad, Walid}]{chehimi_scaling_2023}
Chehimi M, Pouryousef S, Panigrahy NK, Towsley D, Saad W.: Scaling {Limits} of {Quantum} {Repeater} {Networks}.
\newblock arXiv; 2023.
\newblock \urlprefix\url{http://arxiv.org/abs/2305.08696}, arXiv:2305.08696 [cs].

\bibitem[{Deutsch et~al.(1996)Deutsch, D. and Ekert, A. and Jozsa, R. and Macchiavello, C. and Popescu, S. and Sanpera, A.}]{deutsch_quantum_1996}
Deutsch D, Ekert A, Jozsa R, Macchiavello C, Popescu S, Sanpera A.
\newblock Quantum privacy amplification and the security of quantum cryptography over noisy channels.
\newblock Physical Review Letters. 1996 Sep;77(13):2818--2821.
\newblock \urlprefix\url{http://arxiv.org/abs/quant-ph/9604039}, arXiv:quant-ph/9604039, \doi{10.1103/PhysRevLett.77.2818}.

\bibitem[{Dür and Briegel(2007)Dür, W. and Briegel, H. J.}]{dur_entanglement_2007}
Dür W, Briegel HJ.
\newblock Entanglement purification and quantum error correction.
\newblock Reports on Progress in Physics. 2007 Aug;70(8):1381--1424.
\newblock \urlprefix\url{http://arxiv.org/abs/0705.4165}, arXiv:0705.4165 [quant-ph], \doi{10.1088/0034-4885/70/8/R03}.

\bibitem[{Dür et~al.(1999)Dür, W. and Briegel, H.-J. and Cirac, J. I. and Zoller, P.}]{dur_quantum_1999}
Dür W, Briegel HJ, Cirac JI, Zoller P.
\newblock Quantum repeaters based on entanglement purification.
\newblock Physical Review A. 1999 Jan;59(1):169--181.
\newblock \urlprefix\url{http://arxiv.org/abs/quant-ph/9808065}, arXiv:quant-ph/9808065, \doi{10.1103/PhysRevA.59.169}.

\bibitem[{Guha et~al.(2015)Guha, Saikat and Krovi, Hari and Fuchs, Christopher A. and Dutton, Zachary and Slater, Joshua A. and Simon, Christoph and Tittel, Wolfgang}]{guha_rate-loss_2015}
Guha S, Krovi H, Fuchs CA, Dutton Z, Slater JA, Simon C, et~al.
\newblock Rate-loss analysis of an efficient quantum repeater architecture.
\newblock Physical Review A. 2015 Aug;92(2):022357.
\newblock \urlprefix\url{https://link.aps.org/doi/10.1103/PhysRevA.92.022357}, \doi{10.1103/PhysRevA.92.022357}.

\bibitem[{Haldar et~al.(2024)Haldar, Stav and Barge, Pratik J. and Cheng, Xiang and Chang, Kai-Chi and Kirby, Brian T. and Khatri, Sumeet and Wong, Chee Wei and Lee, Hwang}]{haldar_reducing_2024}
Haldar S, Barge PJ, Cheng X, Chang KC, Kirby BT, Khatri S, et~al.: Reducing classical communication costs in multiplexed quantum repeaters using hardware-aware quasi-local policies.
\newblock arXiv; 2024.
\newblock \urlprefix\url{http://arxiv.org/abs/2401.13168}, arXiv:2401.13168 [quant-ph].

\bibitem[{Iñesta et~al.(2023)Iñesta, Álvaro G. and Vardoyan, Gayane and Scavuzzo, Lara and Wehner, Stephanie}]{inesta_optimal_2023}
Iñesta ÃG, Vardoyan G, Scavuzzo L, Wehner S.
\newblock Optimal entanglement distribution policies in homogeneous repeater chains with cutoffs.
\newblock npj Quantum Information. 2023 May;9(1):1--7.
\newblock \urlprefix\url{https://www.nature.com/articles/s41534-023-00713-9}, publisher: Nature Publishing Group, \doi{10.1038/s41534-023-00713-9}.

\bibitem[{Laurenza et~al.(2022)Laurenza, Riccardo and Walk, Nathan and Eisert, Jens and Pirandola, Stefano}]{laurenza_rate_2022}
Laurenza R, Walk N, Eisert J, Pirandola S.
\newblock Rate limits in quantum networks with lossy repeaters.
\newblock Physical Review Research. 2022 May;4(2):023158.
\newblock \urlprefix\url{https://link.aps.org/doi/10.1103/PhysRevResearch.4.023158}, \doi{10.1103/PhysRevResearch.4.023158}.

\bibitem[{Mantri et~al.(2024)Mantri, Prateek and Goodenough, Kenneth and Towsley, Don}]{mantri_comparing_2024}
Mantri P, Goodenough K, Towsley D.: Comparing {One}- and {Two}-way {Quantum} {Repeater} {Architectures}; 2024.
\newblock \urlprefix\url{http://arxiv.org/abs/2409.06152}, arXiv:2409.06152 [quant-ph].

\bibitem[{Muralidharan et~al.(2016)Muralidharan, Sreraman and Li, Linshu and Kim, Jungsang and Lütkenhaus, Norbert and Lukin, Mikhail D. and Jiang, Liang}]{muralidharan_optimal_2016}
Muralidharan S, Li L, Kim J, Lütkenhaus N, Lukin MD, Jiang L.
\newblock Optimal architectures for long distance quantum communication.
\newblock Scientific Reports. 2016 Feb;6:20463.
\newblock \doi{10.1038/srep20463}.

\bibitem[{Pirandola(2019)Pirandola, Stefano}]{pirandola_end--end_2019}
Pirandola S.
\newblock End-to-end capacities of a quantum communication network.
\newblock Communications Physics. 2019 May;2(1):1--10.
\newblock \urlprefix\url{https://www.nature.com/articles/s42005-019-0147-3}, publisher: Nature Publishing Group, \doi{10.1038/s42005-019-0147-3}.

\bibitem[{Pirandola et~al.(2017)Pirandola, Stefano and Laurenza, Riccardo and Ottaviani, Carlo and Banchi, Leonardo}]{pirandola_fundamental_2017}
Pirandola S, Laurenza R, Ottaviani C, Banchi L.
\newblock Fundamental {Limits} of {Repeaterless} {Quantum} {Communications}.
\newblock Nature Communications. 2017 Apr;8(1):15043.
\newblock \urlprefix\url{http://arxiv.org/abs/1510.08863}, arXiv:1510.08863 [quant-ph], \doi{10.1038/ncomms15043}.

\bibitem[{Reichardt et~al.(2024)Reichardt, Ben W. and Aasen, David and Chao, Rui and Chernoguzov, Alex and van Dam, Wim and Gaebler, John P. and Gresh, Dan and Lucchetti, Dominic and Mills, Michael and Moses, Steven A. and Neyenhuis, Brian and Paetznick, Adam and Paz, Andres and Siegfried, Peter E. and da Silva, Marcus P. and Svore, Krysta M. and Wang, Zhenghan and Zanner, Matt}]{reichardt_demonstration_2024}
Reichardt BW, Aasen D, Chao R, Chernoguzov A, van Dam W, Gaebler JP, et~al.: Demonstration of quantum computation and error correction with a tesseract code.
\newblock arXiv; 2024.
\newblock \urlprefix\url{http://arxiv.org/abs/2409.04628}, arXiv:2409.04628 [quant-ph].

\bibitem[{Shor(1995)Shor, Peter W.}]{shor_scheme_1995}
Shor PW.
\newblock Scheme for reducing decoherence in quantum computer memory.
\newblock Physical Review A. 1995 Oct;52(4):R2493--R2496.
\newblock \urlprefix\url{https://link.aps.org/doi/10.1103/PhysRevA.52.R2493}, \doi{10.1103/PhysRevA.52.R2493}.

\bibitem[{Steane(1996)Steane, A. M.}]{steane_error_1996-1}
Steane AM.
\newblock Error {Correcting} {Codes} in {Quantum} {Theory}.
\newblock Physical Review Letters. 1996 Jul;77(5):793--797.
\newblock \urlprefix\url{https://link.aps.org/doi/10.1103/PhysRevLett.77.793}, \doi{10.1103/PhysRevLett.77.793}.

\bibitem[{Wootters and Zurek(1982)Wootters, W. K. and Zurek, W. H.}]{wootters_single_1982}
Wootters WK, Zurek WH.
\newblock A single quantum cannot be cloned.
\newblock Nature. 1982 Oct;299(5886):802--803.
\newblock \urlprefix\url{https://www.nature.com/articles/299802a0}, publisher: Nature Publishing Group, \doi{10.1038/299802a0}.

\bibitem[{Yakar and Ben-Or(2025)Yakar, I. and Ben-Or, M.}]{global_epp_results}
Yakar I, Ben-Or M.: Interactive Web Visualizations for Distillation Policy Results; 2025.
\newblock \urlprefix\url{https://deduckproject.github.io/global-epp-results}, accessed: 2025-08-01.
\newblock \url{https://doi.org/10.5281/zenodo.17292576}.

\bibitem[{Żukowski et~al.(1993)Żukowski, M. and Zeilinger, A. and Horne, M. A. and Ekert, A. K.}]{zukowski_event-ready-detectors_1993}
Żukowski M, Zeilinger A, Horne MA, Ekert AK.
\newblock ``{Event}-ready-detectors'' {Bell} experiment via entanglement swapping.
\newblock Physical Review Letters. 1993 Dec;71(26):4287--4290.
\newblock \urlprefix\url{https://link.aps.org/doi/10.1103/PhysRevLett.71.4287}, publisher: American Physical Society, \doi{10.1103/PhysRevLett.71.4287}.

\end{thebibliography}

\end{document}